# The new science of metagenomics and the challenges of its use in both developed and developing countries

Edi Prifti[1] & Jean-Daniel Zucker[2,3,4]

[1]Institut National de la Recherche Agronomique, US 1367 MGP, 78350 Jouy-en-Josas, France;

[2]Institut de Recherche pour le Développement, UMI 209, Unité de modélisation mathématique et informatique des Systèmes Complexes, 93143, Bondy, France;

[3]Institut National de la Santé et de la Recherche Médicale, U 872, Nutriomique, Équipe 7, Centre de Recherches des Cordeliers, 75006, Paris, France;

[4]Vietnam National University, Equipe MSI, Institut de la Francophonie pour l'Informatique, Hanoi, Vietnam;

**Abstract**

*Our view of the microbial world and its impact on human health is changing radically with the ability to sequence uncultured or unculturable microbes sampled directly from their habitats, ability made possible by fast and cheap next generation sequencing technologies. Such recent developments represents a paradigmatic shift in the analysis of habitat biodiversity, be it the human, soil or ocean microbiome. We review here some research examples and results that indicate the importance of the microbiome in our lives and then discus some of the challenges faced by metagenomic experiments and the subsequent analysis of the generated data. We then analyze the economic and social impact on genomic-medicine and research in both developing and developed countries. We support the idea that there are significant benefits in building capacities for developing high-level scientific research in metagenomics in developing countries. Indeed, the notion that developing countries should wait for developed countries to make advances in science and technology that they later import at great cost has recently been challenged.*

## Introduction

Our view of the microbial world and its impact on our lives is rapidly changing. Until recently we have considered ourselves as largely independent from the microbial ecosystem we live in (Blaser 2006; Ley, Lozupone et al. 2008; Davies 2009). The mainstream thinking that we would be healthier when staying away from microbes is now debated (Bloomfield, Stanwell-Smith et al. 2006); hygiene and antibiotics excess during childhood may be associated with allergies and asthma during adulthood (Hanski, von Hertzen et al. 2012; Kawamoto, Tran et al. 2012; Russell, Gold et al. 2012).

Bacteria, one of the first organisms on Earth (present for more than 3 billion years), have been evolving and adapting to all sorts of environments ever since, elaborating a large genetic pool that codes for many biological pathways that perform a plethora of functions (Canganella and Wiegel 2011). In the interconnected web of life where microbes are most abundant (Whitman, Coleman et al. 1998), humans, like other multicellular organisms, have evolved to live in equilibrium and in symbiosis with them. Indeed, our genome does not code for all the biological functions needed for our survival, or to take full advantage of the environment we live in. We interact considerably with our microbiota and as a consequence the health of this ecosystem is tightly linked to ours.

Most abundant in the intestine, the gut microbiota is now considered to be an organ reaching approximately 2kg in mass (Baquero and Nombela 2012). With 150 times more genes than our own genome, the collective genome of our microbiome (also called "our other genome" or metagenome) codes for many different functions that are not undertaken by our cells. For instance, gut bacteria can protect us by producing anti-inflammatory factors, antioxidants and vitamins, but also harm us by producing toxins that mutate DNA, or affecting the nervous and immune systems. The outcome of microbiome deregulation may take the form of various chronic diseases, including obesity, diabetes and even cancers (Zhao 2010). The very nature of human identity is now being questioned and an increasing number of scientists believe that we are indeed a super-organism with a microbial majority: 10 times more microbes than human cells, that should be taken into consideration as part of us (Blaser 2006; Gill, Pop et al. 2006; Davies 2009).

Gene therapy was developed at the beginning of the twenty-first century and came with the promise of revolutionizing medicine, but its implementation was more challenging than anticipated. Such difficulties were partially due to the multifactorial nature of most diseases but also to the complex implementation and success rate of such therapy. The gut microbiota opens new means of intervention in curing complex diseases linked to it, such as the use of probiotics, fecal transplantation or other microbiome targeted approaches (Borody and Khoruts 2012; Lemon, Armitage et al. 2012; Shanahan 2012). Such interventions are thought to be simpler than any human gene therapy and are of great economical potential for both the private and public health sector.

Prokaryotes are some of the most diverse organisms on the planet bearing many known and unknown functions that affect nearly all aspects of life on Earth. For instance, the bacteria that populate the ocean affect key chemical balances in the atmosphere and ensure the very habitability of the Earth. Also it is a known fact that soil microorganisms are fundamental for terrestrial processes as they play an important role in various biogeochemical cycles by contributing to plant nutrition and soil health (Mocali and Benedetti 2010). As such microorganisms are of great hope for scientific research and potential biotechnological applications. This increasing interest in understanding the role of the microbiome in planet ecology, health and disease as well as other biotechnological applications promises very important economic and societal benefits for those countries that are involved in such research. The holistic study of the human microbiome is a fairly new approach that necessitates very expensive cutting-edge technologies and multi-disciplinary teams. Only big research institutions with large funding programs, usually from developed countries, are currently able to undertake such projects, leaving developing countries behind in this field.

This chapter introduces first the new science of metagenomics and its many challenges while reviewing some of the major discoveries up to date. We discuss next the benefits that developing countries might reap if they were to build the needed infrastructure and become involved in microbiome research.

## The new science of Metagenomics

Now that we have established the importance of the microorganisms that live inside and around us, let us focus on the available methods and tools used to study them. An estimated 99% of the prokaryotes are difficult to study in isolation for several reasons (Streit and Schmitz 2004; Schloss and Handelsman 2005): (i) they depend on other organisms for critical processes, (ii) fail to grow *in vitro* or (iii) have even become extinct in fossil records (Tringe and Rubin 2005). These obstacles can be bypassed by focusing on DNA, a very stable molecule that can be isolated directly either from living or dead cells. Usually DNA extracted from a given sample belongs to different microbial genomes constituting what is termed a metagenome. The study of metagenomes is a new emerging field, which is referred to as metagenomics (NRCC 2007). The progressive reduction in the cost of high-throughput sequencing made possible to sequence large quantities of DNA from mixtures of organisms (Shendure, Mitra et al. 2004; Metzker 2010) offering a very detailed insight into entire ecosystems previously thought to be inaccessible.

Quantitative metagenomics focuses on quantifying DNA molecules in a given sample as opposed to functional metagenomics which focuses on clone expression (Lakhdari, Cultrone et al. 2010). Quantitative metagenomics, on which we will mostly focus in this chapter, can be approached through different strategies (Gabor, Liebeton et al. 2007). The sequencing of the 16S ribosomal RNA gene is one of the most accessible and thus most frequently used approaches in quantitative metagenomics. Prior studies of bacterial evolution and phylogenetics provided the foundation for subsequent applications of sequencing based on 16S rRNA genes for microbial identification (Winker and Woese 1991). Indeed the 16S rRNA genes consist of highly conserved region sequences that alternate with regions of variable nucleotide sequence, which are used for taxonomic classification. The16S rRNA gene is, thus, a good marker to explore the phylogenetic composition of a given sample, identify new species or even unknown phylogenetic groups. In quantitative metagenomics variable regions of bacterial 16S rRNA genes are usually amplified by PCR and then subjected to library construction followed by sequencing using next-generation technologies. The pool of sequenced reads are then clustered, may be mapped onto a database of previously characterized sequences, and used for further analyses in the studied context. Microbial 16S rDNA sequencing is considered the gold standard for characterizing microbial communities, but this approach would fail to capture information about what the functions of different organisms are, knowing that organisms with identical 16S sequencing may perform very different functions. A good example is the difference between various strains of *Escherichia coli* (*Enterohaemoragic – EHEC, Enterotoxic – ETEC, Enteroaggregative - EAEC*) and related organisms such as *Shigella sonnei*, which have different clinical manifestations, different treatment modalities, yet are undistinguishable by 16S rRNA sequences (Harris and Hartley 2003).

To overcome the limitations of the 16S rRNA profiling approach, the sequencing of entire microbial genomes (made possible by next generation sequencing technologies) constitutes a very attractive strategy for comprehensive metagenomics studies. The whole-genome (also called WMS for whole metagenomic sequencing) approach is increasingly used and has already produced many interesting results that we discuss in the next two sections. Figure 1 illustrates an overview of a WMS pipeline from the collection of samples to the generation of hypotheses and the testing of prediction models. Sample collection from a given environment is a crucial process since the microbial communities may be quite different between two very close locations (as is the case, for example, for soil environments) and should be determined according to the project needs. The DNA extraction protocols are also deciding factors and depend on the microbial composition of the sample. For example Gram-positive bacteria, which are hard to lyse organisms, might be underrepresented or overrepresented in environmental DNA preparations depending on the extraction protocol. Sequencing followed by mapping onto a selected reference gene catalog and bioinformatics pre-treatment analyses constitute another important part of this pipeline that will ensure that the biological signal is isolated while reducing the noise caused by the technical variability throughout the study. Finally, the use of the right statistical tools and datasets will be crucial in hypothesis generation and testing. We discuss later in this chapter the different issues and challenges faced at each stage of this process (cf. Figure 1).

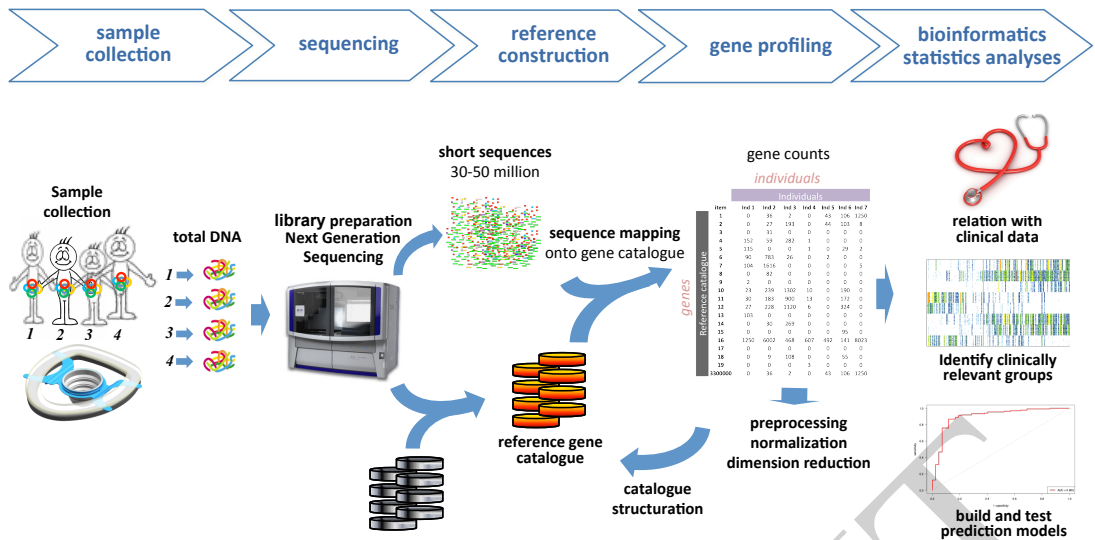

*Figure 1: Overview of a whole-metagenome-sequencing project from sample collection to hypotheses generation (after N. Pons & E. Le Chatelier).*

Another important and increasingly used application of WMS is the study of gene expression. The sequencing of cDNA, which corresponds to the whole RNA in a given sample, has brought many new application possibilities to scientists. With cDNA microarrays, a gene expression measuring technology, it is possible to focus only on those transcripts that have a corresponding probe on the chip and which are usually linked to coding sequences. RNA-Seq technology allows to bypass this limitation and gives a true holistic view of the transcriptome (Shendure 2008). Largely used in single genome transcriptomics it is now starting to be applied in metatranscriptomics settings. The RNA-Seq approach offers an unprecedented resolution on both the activity of a given bacteria and the functional dynamics of the genes. On the other hand, it comes with a price to pay, that of the analytical challenge that underlies the complexity behind the very large number of variables in the data. Other "meta-omics" approaches such as metaproteomics or meta-metabolomics are still in their infancy but just as promising.

The precise bio-characterization of samples from different environments of interest is increasingly becoming a routine with the help of metagenomics and other meta-omics technologies and this new science is advancing very quickly. Many discoveries are being made in relation to human health and the environment as we discuss hereafter.

*Metagenomics in health and disease*

Many projects have been funded these last years aiming to characterize the human microbiome and uncover its impact in human health and disease. One of the first internationally coordinated efforts was the European-funded MetaHIT[1] project (Ehrlich and MetaHIT 2010), which started in early 2008. Its main objective was to establish associations between the genes of the human intestinal microbiota and health and disease, focusing on two main disorders of increasing importance in Europe: Inflammatory Bowel Disease (IBD) and obesity. One of the first important achievements was the establishment of an extensive reference catalog of microbial genes present in the human intestine. Indeed, more than 85% of gut bacteria are unknown, and more than 80% of them are considered today unculturable (Eckburg, Bik et al. 2005; Qin, Li et al. 2010). This study offered the first high-resolution picture of the immense diversity and complexity of the gut microbiota. The size of our intestinal metagenome is 150 times larger than that of our own genome and is constituted of more than three million non-redundant microbial genes, which are largely shared among the individuals of the studied cohort. Over 99% of them are bacterial genes, indicating that the entire cohort harbors more than 1,000 prevalent bacterial species and each individual at least 160 such species (Qin, Li et al. 2010).

The HMP[2] is another major project funded by the NIH. Its main goals are to characterize the microbial communities found at several different sites on the human body, including nasal passages, oral cavities, skin, the gastrointestinal and urogenital tracts, and to analyze the role of these microbes in human health and disease (Group, Peterson et al. 2009). They found that, in a cohort of healthy people, oral and stool communities were especially diverse in terms of community membership, while vaginal sites harbored particularly simple communities. Additionally, even though the diversity and abundance signature of each body site were found to vary among individuals, a niche specialization as well as the metagenomic carriage of metabolic pathways were observed to be stable among the subjects (Human Microbiome Project 2012). Overall, in healthy humans, microbiota tend to occupy a range of distinct configurations from many of the disease-related perturbations studied to date (Sokol, Pigneur et al. 2008; Qin, Li et al. 2010).

As a consequence of the complexity of metagenomics data and the relatively new age of this field, significant effort was needed in building the analytical framework as well as the associated bioinformatics pipelines and tools. Both of the aforementioned projects among other more isolated initiatives, helped in developing such technologies many of the current projects are now using in order to discover associations between the gut microbiome and clinical phenotypes and diseases (Ehrlich and MetaHIT 2010; Qin, Li et al. 2010; Human Microbiome Project 2012; Morgan and Huttenhower 2012).

---

[1] Acronym for "Metagenomics of the Human Intestinal Tract". URL: http://www.metahit.eu/
[2] Acronym for "Human Microbiome Project". URL: http://www.hmpdacc.org/

One of the properties of a sampled ecosystem is species diversity, which when highly diverse is usually linked with good health. Indeed, it was discovered that low diversity in gut microbiota is associated with several human diseases such as obesity and inflammatory bowel disease (Turnbaugh, Hamady et al. 2009; Qin, Li et al. 2010). In some other cases high diversity may be associated with disease such as bacterial vaginosis for example (Srinivasan, Hoffman et al. 2012).

A recent study showed the involvement of intestinal flora in type-2-diabetes on a Chinese cohort. Approximately 60,000 microbial genes were found to be differentially abundant among type-2-diabetic patients who were also characterized by a moderate degree of gut microbial dysbiosis, a decrease in the abundance of some universal butyrate-producing bacteria and an increase in various opportunistic pathogens. The authors also demonstrated that these gut microbial markers might be useful for classifying type-2-diabetic patients based only on their fecal samples (Qin, Li et al. 2012).

Intestinal flora was also related to the inflammatory status of the host in symptomatic atherosclerosis patients, who were found to be enriched in the genus *Collinsella* as opposed to the controls enriched in *Eubacterium* and *Roseburia* (Karlsson, Fak et al. 2012). Even though this study cannot provide evidence for direct causal effects, these findings indicate that the gut metagenome may play a role in the development of systematic atherosclerosis knowing that inflammation is an important contributor to the pathogenesis of atherosclerosis (Hansson 2005).

Accumulating evidence now indicates that the gut microbiota also communicates with the central nervous system, possibly through neural, endocrine and immune pathways, and thereby influences brain function and behavior (Grenham, Clarke et al. 2011; Cryan and Dinan 2012). Studies in germ-free animals and in animals exposed to pathogenic bacterial infections, probiotic bacteria or antibiotic drugs suggest a role for the gut microbiota in the regulation of anxiety, mood, cognition and pain. Factors, including infection, disease and antibiotics, may alter the stability of the natural composition of the gut microbiota and thereby have a deleterious effect on the well-being of the host (Forsythe, Sudo et al. 2010).

Another study demonstrated the key role of the gut microbiota in immuno-modulatory mechanisms underlying multiple sclerosis. Mice genetically predisposed to spontaneously develop EAE (Experimental Autoimmune Encephalomyelitis, an animal model for multiple sclerosis disease) were housed under germ-free conditions and, as a result, remained fully protected from EAE throughout their life until this protection dissipated upon colonization with conventional microbiota in adulthood (Berer, Mues et al. 2011). Several small studies have demonstrated links between altered intestinal microbiota in children with autism as compared with controls (Finegold, Dowd et al. 2010; Adams, Johansen et al. 2011; Finegold, Downes et al. 2012). These relations may be explained however by different factors such as diet and larger controlled clinical studies are needed for more evidence.

These are but a few studies among many others that have taken the first steps in demonstrating the existence associations between intestinal flora and different human diseases. Scientists haven't had enough time yet to gather evidence on causality but this is the next step. Meanwhile there have already been some results indicating how we can use metagenomics and the microbiome to improve our health (Shanahan 2012). A first application area is personalized medicine. Information on the human microbial ecosystems may help stratify individuals and reduce the variability of the cohort so that a given treatment is more effective if adapted to any of the different sub-phenotypes. For instance, an unexpected discovery was the identification of enterotypes, three robust clusters, which remain consistent among different countries and cohorts and are stable over time (Arumugam, Raes et al. 2011). Even though the discrete nature of enterotypes is debated (Jeffery, Claesson et al. 2012), they are found to be associated with long-term diet, particularly protein and animal fat (*Bacteroides*) versus carbohydrates (*Prevotella*) (Wu, Chen et al. 2011).

The potential of the human microbiome as an early detection biomarker for diagnostic and prognostic purposes is a very active area of research. Oral or fecal microbial samples can be obtained very easily and used immediately as diagnostic tools. As an example microbial genes associated with type-2-diabetes (Qin, Li et al. 2012) were used to construct prediction models that could correctly classify a sample with an accuracy of greater than 80%. This is considerably better than models based on the human genes (66%) linked with type-2-diabetes by genome wide association studies (van Hoek, Dehghan et al. 2008). Most clinical indicators are not optimal such as the OGCT[3] for diabetes or BMI for obesity (Romero-Corral, Somers et al. 2008). Newer, more biologically relevant indicators are needed. The microbial biomarkers can help in this quest and could even be used to predict future occurrences of a disease.

The human microbiome and especially the gut flora offer a yet to-be-appreciated potential for interventional medicine and open unprecedented possibilities for curing human diseases. One area of intervention consists in modulating the disrupted microbial ecosystem and bringing it close to a normal state. This can be achieved through different ways such as by using prebiotics and probiotics (Sharp, Achkar et al. 2009; Gareau, Sherman et al. 2010). The use of probiotics has been already shown to be successful in animal studies where *Lactobacilli* and *Bifidobacteria* based probiotics can alleviate visceral pain induced by stress (Verdu, Bercik et al. 2006); the role of probiotics in treating diseases such as IBS has also been shown (Aragon, Graham et al. 2010).

---

[3] Acronym for "Oral Glucose Challenge Test" a score on which diabetes classification and diagnostics is based.

Another way in achieving ecosystem modulation is through fecal transplant, also known as fecal flora reconstitution (Baquero and Nombela 2012; Borody and Khoruts 2012). Different studies have already shown the success of this approach in treating extreme cases of *Clostridium difficile* infections that were resistant to antibiotics (van Nood, Vrieze et al. 2013). Despite the very high success rate (>90% - far better than many drugs), there are still ethical issues that need to be addressed and more clinical studies should be performed for this intervention approach to be more widely accepted and used in the medical field. Finally, therapeutic drugs can be designed to directly interact with the microbiome and modulate its different functions to change its state and thus transform the diseased phenotype into a healthy state (Jia, Li et al. 2008; Haiser and Turnbaugh 2012).

*Environmental metagenomics*

Microorganisms represent the largest reservoir of genetic diversity on Earth, outnumbering all other organisms (NRCC 2007). As an example, bacteria are responsible for about half of the photosynthesis on Earth. In spite of their crucial role, prokaryotic diversity still suffers from one of the greatest knowledge gaps in the biological sciences and remains largely unexplored and unexploited (Rodriguez-Valera 2004). There is no universally agreed estimate about their real total number, their real diversity or what principles govern their origin and change. Some researchers estimate the total number of prokaryotic cells on earth at $5 \times 10^{30}$ including $10^6$-$10^8$ individual genomes belonging to different species (Sleator, Shortall et al. 2008). For the soil, there are estimates ranging between 3,000 and 11,000 microbial genomes per gram of soil (Schmeisser, Steele et al. 2007), which makes it clear that current technologies could not support complete sequencing of such highly diverse environments (Kowalchuk, Speksnijder et al. 2007). Beyond the inter species diversity, there is also an intra species diversity that has been overlooked but that has important consequences. For example, in an easy-to-cultivate species such as *Escherichia coli* may lay a vast gene pool that is not accessible by studying one single strain. Indeed, the diversity of the genes within a bacterial species is another important facet of prokaryotic diversity (Boucher, Nesbo et al. 2001). Metagenomics as a culture-independent genomic analysis can also help discover more about the microbial diversity of natural environments such as soil, water and sediments (Lopez-Garcia and Moreira 2008) and has applications in agriculture, sustainability, engineering and environment.

*Soil*

It is a known fact that soil microorganisms are fundamental for terrestrial processes as they play an important role in various biogeochemical cycles by contributing to plant nutrition and soil health (Mocali and Benedetti 2010). This "hidden" biodiversity could be a great resource of natural products for agricultural and biotechnological applications (Steele and Streit 2005). Assessing and preserving the diversity of soil microorganisms is thus crucial. The most critical biotransformations at stake (degradation of pollutants, synthesis of biofuels and production of novel drugs) require a whole microbial community to be performed. For instance, no single microbe is capable of converting ammonia to nitrate but teams of microbes can do this very efficiently. These communities are likely to explain the farming mystery of "suppressive soil" in which a pathogen is known to persist but causes little damage to the crops. The activities of suppressive soil communities are quite beneficial to agriculture ensuring the quality and provision of ecosystem services (NRCC 2007).

Sequencing the soil metagenome could bring large economic and environmental value but represents a task of unprecedented magnitude. A coordinated international effort was established to combine the skills of the global scientific community to focus on sequencing and annotating the soil metagenome: TerraGenome (Vogel, Simonet et al. 2009). Launched in September 2011, the project relies on many existing bioinformatics resources, RDP (Ribosomal Database Project), QIIME (Quantitative Insights Into Microbial Ecology), Greengenes (a web application providing access to the current and comprehensive 16S rRNA gene sequence alignment for browsing, blasting, probing, and downloading), IMG/M (Integrated Microbial Genomes/Microbiome) to name a few. The complete sequencing of a "reference" soil metagenome is aimed and the agroecology field experiment of Park Grass that has been running for more than 150 years at the UK agricultural sciences institute was chosen for investigation. There are several projects involved in soil sequencing and a good review of soil metagenomics studies and their relevance for biotechnology and ecology is made by Van Elsas et al. (van Elsas, Costa et al. 2008).

*Marine Water*

The marine environment is the largest contiguous ecosystem on Earth, occupying 71% of the earth's surface with an average depth of 4 km (Karl 2007). It is not surprising that the oceans represent one of the most significant yet least understood microbial-driven natural environments on the planet (Martin-Cuadrado, Lopez-Garcia et al. 2007). The bacteria that populate the ocean affect key chemical balances in the atmosphere and play an important role in maintaining the very habitability of the planet.

One of the first studies of marine environmental genomics was led by DeLong. This project was quite successful since the scientists involved in it discovered in archaea and bacteria the presence of a family of genes called rhodopsins. These genes were only known to exist before in plants where they play a role in collecting energy from the sun (Beja, Aravind et al. 2000). Except for some species that use rhodopsins to capture energy, many others use them for communication or environmental awareness. The *Sargasso Sea project* was another one who set a global expedition to gather microbes from the world's oceans and sequence their DNA. The authors found 1,800 species of microbes, including 150 new species of bacteria, and over 1.2 million new genes (Venter, Remington et al. 2004). Unfortunately, the Sargasso Sea Project unraveled only a small amount of the diversity. As Gilbert et al. noted, "*with approximately 1 million bacteria per milliliter of seawater and an estimated average genome size of 2 million bp, the Sargasso Sea project sequenced only 0.05% of the genomic information in a single milliliter - a proverbial drop in the ocean*" (Gilbert and Dupont 2011).

We are only at the beginning of a new vast domain of analysis of the biodiversity that needs to be achieved as soon as possible so that the loss of diversity that is foreseen in all habitats can be assessed as early as possible. Karl et al. review several case studies which illustrate the need for «*comprehensive analyses - ranging from genomes to biomes, coupled to interdisciplinary physical and chemical observations of broad temporal - spatial scales - before a comprehensive understanding of the role of microorganisms in oceanic ecosystems can be achieved* " (Karl 2007). There are many other projects related to metagenomics and marine environment. The Marine Microbiology Initiative (MMI) lunched a project aiming to sequence, assemble, and annotate approximately 200 diverse marine phage/virus genomes and to sequence approximately 50 viral metagenomes from an array of marine environments.

These studies, be it for health or environmental applications, are just the tip of the iceberg and many are yet to come and bring fascinating and unimagined insight on the importance of the microbiome in our lives. This comes as a result of the evolution of genomics into metagenomics. The field is in its infancy and many challenges are yet to be addressed as discussed in the next section of this chapter.

## Quantitative metagenomics and its challenges

As for every new field, many tools and approaches need to be invented from scratch or adapted in order to answer questions and gain insight on the studied topic, and metagenomics is no exception. It is overwhelming to see the pace at which the whole experimental and analytical framework is being built as a consequence of international teamwork and different collaborations. Here we discuss some of the challenges that still need to be addressed in order to improve the efficacy of future investigations.

*Experimental challenges and protocols*

The classical whole-metagenome-sequencing pipeline as illustrated in Figure 1 is composed of many steps, several of which are related to the experimental biology. The first step after experimental design, which we will not discuss here, is the sample collection. This component is crucial since samples contain the biology we are trying to study and understand. Protocols should be precise and applied similarly throughout the samples, but this is easier said than done since in some cases (such as in the study of human gut microbiome), fecal samples are extracted and stored by patients themselves. Storage is very important (Wallenius, Rita et al. 2010) and depends on the studied ecosystem but might be quite challenging when sampling, for example, from deep sub-seafloor sediments (Alain, Callac et al. 2011) or other extreme environments. One key aspect specific to WMS strategies is the requirement for ever-greater amounts of input genomic DNA for comprehensive metagenomics studies (Petrosino, Highlander et al. 2009). This is an important limitation when the starting material is limited, as in paleogenomics (Tringe and Rubin 2005).

Another crucial issue is the process of DNA extraction. By definition, in a microbial community there are many different species and phylogenetic groups and as a consequence the DNA is encapsulated in cells with different properties. The techniques that are used to lyse cells might also affect the composition of environmental DNA libraries, as the harsh lysis methods that are necessary to extract DNA from every organism will cause degradation of the DNA from some organisms (Tringe and Rubin 2005). Hard-to-lyse cells, such as Gram-positive bacteria, might therefore be underrepresented or overrepresented in environmental DNA preparations.

To coordinate the activities carried out within different programs, the International Human Microbiome Consortium (IHMC)[4], has been constituted and formally announced in October 2008. This has facilitated identification of a clear need to standardize[5] the procedures in the Human Microbiome research. The overall concept of IHMS is to promote the development and implementation of standard procedures and protocols in three separate but related fields: (i) collecting and processing of human samples, (ii) sequencing of the human-associated microbial genes and genomes and (iii) organizing and analyzing the data gathered.

*Bioinformatics/biostatistics challenges*

---

[4] http://www.humanmicrobiome.org/
[5] http://www.microbiome-standards.org/

Classical techniques in biology have been around for some time now; they are well tuned and have proved to be effective over the years. Solving some of the issues mentioned above and implementing standard protocols would greatly benefit in reducing experimental variability. The problem, however, is more complicated from an analytical point of view. Indeed, the amount of information generated by metagenomic studies is unprecedented and in addition to the informatics infrastructure needed to deal with the data deluge (Pennisi 2011), bioinformatics and biostatistics challenges need to be addressed promptly.

One of the main challenging issues concerns the gene reference catalogue. The main difficulty is that the gene reference catalogue needs to be representative enough for each studied sample. Reducing the immense diversity of the microbial word to a single reference catalogue (similar to that of MetaHIT or the Human Genome Project) is still a challenge (Streit and Schmitz 2004). It would contain an astronomical number of genes that we would not be able to practically analyze. The solution for now is to focus on subsets of reference catalogues representative for a given ecosystem such as that of the human gut microbiota (Qin, Li et al. 2010). This might be population-dependent and meta-analyses studies can become challenging.

Gene profiling and data pre-treatment constitute two major research points in the bioinformatics processing of metagenomics data. Gene profiling consists in mapping reads onto the reference catalogue and counting how many of them correspond to each gene. This is far from being a trivial task, and questions such as how to count those reads that map to more genes simultaneously are still a challenge. Similarly, existing pre-processing techniques, like the process that allows transforming the data in order to reduce the technical and experimental noise through normalization techniques, are not adapted to metagenomics data. Most of the genes differ in terms of abundance between different samples, in contrast to gene expression levels in microarray experiments where most of the normalization techniques were introduced (Smyth and Speed 2003).

Finally, one of the biggest challenges that we face is the lack of an adapted statistical framework. The immense dimensionality of the data with millions of variables along with the very particular sparse nature of such data (due to the absence of species and thus genes among samples), make the use of classical statistics unsuitable. Even after log-transforming these data show particularly long-tailed distributions where the use of traditional parametric statistics is not possible. Another major issue beside the difficulty of applying variable selection in such datasets is the strong interdependence that exists among genes (genes of the same genome show very similar abundance profiles) and even among species. Fortunately, some work in this direction has been done and other work is ongoing. Reducing dimensions is one way to tackle the problem, by clustering genes into groups of genes that would correspond to the core-genomes using different techniques such as abundance profiles or gene assembly (Nagarajan and Pop 2013). By identifying bags of genes, a single abundance profile can be calculated and thus the data can be reduced by more than three orders of magnitude and classical statistics could then be applied. Another issue with metagenomics data is that microorganisms very actively exchange parts of their genomes helped sometimes by other smaller genomic entities such as bacteriophages, which render the picture even more complex. This is still the beginning and an increasingly number of statisticians are now reflecting on these issues and proposing solutions.

# The importance of expanding metagenomics in both developed and developing countries

*The economic, industrial and technical promises of metagenomics*

The previous sections of this chapter demonstrate how metagenomics is prefiguring a paradigmatic change in many scientific fields. In fact, this very recent technology is already challenging the very basis of scientific practice in molecular biology. Some authors have even stated that "*the hypothesis-driven science may find it hard to keep up*" (Gilbert and Dupont 2011). There is already a vast number of "habitats" that have been sampled in the past ten years: marine viral community, human feces viral community, drinking water, Sargasso sea, Eel river sediments, farm soil (Minnesota), human gut microbiome (various countries), mammoth fossil, Mediterranean sea, coral reef, etc. (Hugenholtz and Tyson 2008). As indicated earlier, besides for purely environmental studies, metagenomics is becoming more and more important in health and medicine. There are yet unseen limitations to its potential applications ranging from medicine, farming and energy to more generally everything that affects our planet (Lorenz and Eck 2005; Tripathi, Tripathi et al. 2007; Ehrlich and MetaHIT 2010).

In spite of the scientific and industrial excitement about the new possibilities of unraveling microbial diversity, a remaining challenge is to develop technologies at competitive prices and to turn metagenomics into commercial successes. As Lorenz and Eck put it, "*Metagenomics, together with in vitro evolution and high-throughput screening technologies, provides industry with an unprecedented chance to bring biomolecules into industrial application*" (Lorenz and Eck 2005). Both the cost per sequenced base as well as the limitation in human and computational resources to analyze metagenomics data constitute clear bottlenecks. With the appearance of new Next Generation Sequencers, the dye-terminator technology is becoming obsolete and the price per base has dropped by three orders of magnitude (Liu, Li et al. 2012). Moreover, the reduction of sequencing costs comes along with an increase in data quantity, which, without concerted efforts, will quickly outpace the ability of scientists to analyze it (Editorial 2009). As for the resources and costs for microbial sequence analysis they are still extremely expensive although cost-efficient bioinformatics alternative to local computing centers exist (Angiuoli, White et al. 2011).

### *Developing countries investing in metagenomics: from an apparent paradox to a long-term necessity?*

The aforementioned technology and experiments have given to wealthy nations a vast library of data and tools to analyze them, with a great potential for applications and discoveries. Since the beginning of genomics, many developing countries did not participate in the global initiatives for sequencing organisms and, more generally, in genomic research because of the cost of sequencers and lack of qualified staff. At first glance, it might not be so obvious that developing countries should now massively invest in metagenomics given these same reasons and the fact that many other key investments for health and economic development are more urgent. Moreover, for those developing countries who did manage to sequence new organisms it was in most cases with little economic value to them (Coloma and Harris 2009). Several authors have thus argued that developing countries are the ones that can least afford to waste their limited health resources on ineffective diagnostics and therapies, knowing the urgent needs they have to fight infectious diseases and malnutrition (Auffray, Chen et al. 2009). Nevertheless some authors like Séguin et al. (Seguin, Hardy et al. 2008) who even recently argued against such costly investments now challenge the very notion "*that developing countries should wait for developed countries to make advances in science and technology that they later import at great cost*". Because of a slow progress in the areas of scientific research, along with low levels of available funding and investment in sciences in most developing countries, there has been very little scientific contribution toward solving major problems that hinder their global development.

There is no doubt that there is a need for a strong involvement of the scientific community from the developed world, but without taking into account microbial ecosystems from developing countries, any scientific program started at the world scale could not be complete. In other words, developed countries have a great interest in accessing the bio-diversity of developing countries from both scientific and application perspectives. In turn, developing countries that cannot afford expensive next generation sequencers (NGS) may provide some limited and controlled access to their microbiome diversity because they possess unique ecosystems from which valuable knowledge can be learned (Coloma and Harris 2009). In the past there have been bad experiences with "safari research" in which biological samples were taken out of the country for research and did not benefit local populations at all. This kind of behavior has risen concerns in several developing countries and prompted them to pass legislation regarding "*sovereignty over genomics material and data*"[6]. Three countries - Mexico, India and Thailand - have used this concept to express their wish "*to capture the value of their investments*" in large-scale genotyping projects (Virgin and Todd 2011). The positive benefit of legislation associated to this issue is that it can support setting rules for "trading" access to biodiversity with an access to genomics research capabilities. This may be critical for poor countries that do not have yet their own genomics initiatives.

## *Setting priorities for developing countries?*

There are several promising directions of research related to metagenomics in and with developing countries that are worth investigating. We list here a few of them inspired by Djikeng et al.'s seminal paper on this subject (Djikeng, Nelson et al. 2011):

a) *Global surveillance of emerging and reemerging infectious diseases*. Infectious outbreaks in the human population occur regularly in the developing world (Africa, Southeast Asia, and South America) and thus the development of better control measures is crucially important. Early detection and genetic identification of known and unknown pathogens are among the main challenges. Human microbiota studies with the use of metagenomics technology may contribute significantly to discovering emerging pathogens and thus preventing epidemic outbreaks in the developing world.

b) *Understanding sexually transmitted diseases*. There are a growing number of studies (Atashili, Poole et al. 2008) aimed at analyzing the microbial populations in the vaginal ecosystem and how they vary under health and disease conditions. Metagenomics can help identify new clinical biomarkers based on the increasing knowledge on vaginal flora and open new ways of diagnosing and treating STDs such as HIV, which is one of the biggest killers in the developing world.

c) *Enhancing malaria treatment*. Malaria, as one of the most devastating diseases known, mostly impacts developing countries (90% of all malaria deaths occur in sub-Saharan Africa). Metagenomics along with modeling approaches could help monitor the impact of the *Plasmodium* parasite on the microbial communities that reside on and in the human body. This would open new treatment perspectives such as bringing the protective microbiota ecosystem in a normal state with the help of new yet-to-be-developed probiotics (Djikeng, Nelson et al. 2011).

---

[6] A historic protocol was recently initiated at the Nagoya conference on the Convention for biological diversity, called the *Nagoya Protocol on Access to Genetic Resources and the Fair and Equitable Sharing of Benefits Arising from their Utilization to the Convention on Biological Diversity*. (http://www.cbd.int/abs/)

d) *Hunger and malnutrition*. Nutrigenomics focuses on identifying and understanding molecular-level interactions between nutrients and other dietary bioactives with the human genome (Kaput, Ordovas et al. 2005; Ahmed, Haque et al. 2009; Godard and Hurlimann 2009). Metagenomics in the context of nutrigenomics is a unique opportunity to assess the unacceptable consequences of hunger and malnutrition, and to define research priorities that could benefit both developed and developing countries. (Godard and Hurlimann 2009).

e) *Developing new drugs*. Pharmacogenomics, which develop rational means to optimize drug therapy, with respect to the patients' genome and metagenome, to ensure maximum efficacy with minimal adverse effects is now becoming more and more critical to the development of drugs (Jia, Li et al. 2008). Pharmacogenomics may help in the development of research and development capacities which would, in turn, enable developing countries themselves to make meaningful contributions to the advancement of the field (Haiser and Turnbaugh 2012).

f) *Food protection strategies*. Many developing countries are fragile in terms of food security and any plant disease could devastate plantations and cause famine, bringing the country to its knees. Metagenomics research applied to agriculture and soils (Roossinck 2012) may help characterize their microbial diversity and propose new protection strategies.

This list is not exhaustive but it contributes to our argument that it is critical for developing countries to participate to the metagenomics race (Auffray, Chen et al. 2009), in particular because of the applications to genomic medicine by implementing molecular diagnostics and molecular epidemiology. Science and technology, in particular the life sciences, are increasingly recognized as vital components for national progress in developing countries (Virgin and Todd 2011). It is clear that breaking the cycle of dependence of emerging economies towards other developed countries is critical. The costs of healthcare are increasing at a very high rate and relying on developed countries to perform all the molecular diagnostics is not only a financial burden but slows down the building of capacities and the ability to internalize new technologies. Investing in the field of metagenomics and building life sciences-based capacities in developing countries might contribute to improving local health but also potentially stimulate economic growth.

## *The need to increase capacity building in developing countries*

It is not simple to assess the level of technology, resources, and capacity to participate in genomics research for both developing and developed counties. India, Thailand, South Africa,

Indonesia, Brazil, and Mexico, for example, have devoted considerable resources to large-scale population genotyping projects that explore human genetic variation (Coloma and Harris 2009). However, the poorer countries still do not have the resources to develop their own genomic projects on a large scale. Figure 2-4 give an indicator of the importance of metagenomics capacity worldwide. The numbers are self-reported and only give an estimation of the capacities but it is clear that developing countries lag far behind in terms of equipment. For developing countries, building such centers and research facilities to perform metagenomics analysis of their populations is critical for future healthcare infrastructure reducing the gap between the developed and underdeveloped nations.

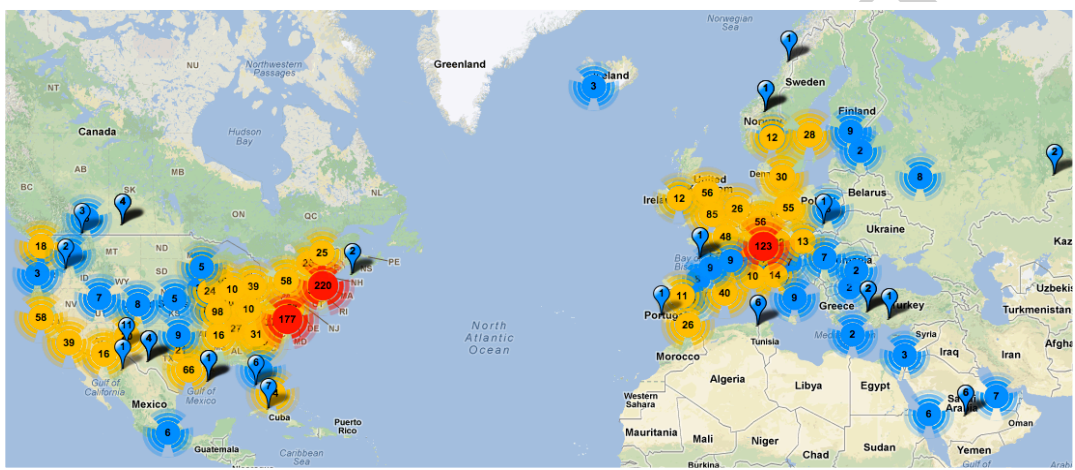

*Figure 2: Estimated map of reported high-throughput sequencers in Western World. The flags represent the research centers and the dots a collection of centers. The numbers indicate the number of sequencers. Their color represents the density from low (blue) to high (red). (OmicsMaps.com data supplied with permission of James Hadfield and Nick Loman.)*

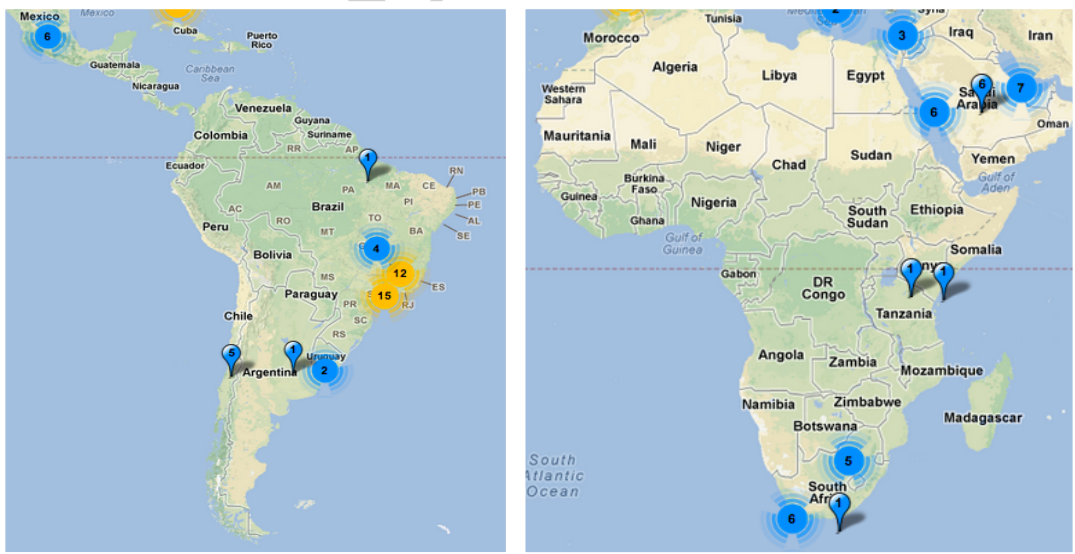

*Figure 3: Estimated map of reported high-throughput sequencers in South-America and Africa. (OmicsMaps.com data supplied with permission of James Hadfield and Nick Loman).*

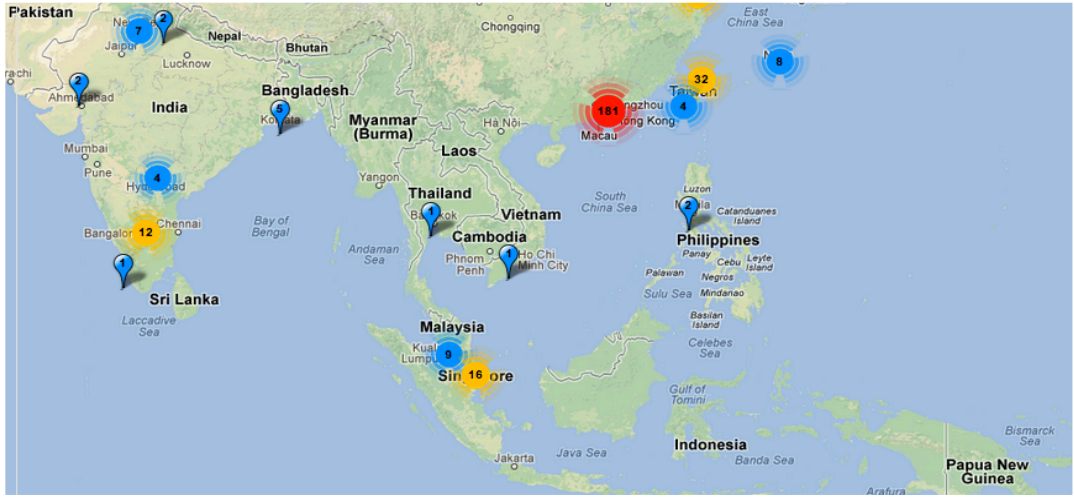

*Figure 4: Estimated map of reported high-throughput South-East Asia. (OmicsMaps.com data supplied with permission of James Hadfield and Nick Loman).*

Sequencing capacity and equipment is not the only key to the problem. Staff, core-facilities, project funding, are of uttermost importance to the development of the research as well. Efforts must be made by developed countries to host and form more students from the South, so that they may turn and invest their knowledge in their countries. It does not always work this way as many students from developing countries choose not to go back and take advantage of more developed research infrastructure in the host countries where they will build their careers. China, for instance, has developed very attractive programs to make sure its PhD students come back. Others like Vietnam fund PhD programs that require working for a few years in their university after completion.

A second bottleneck for the development of metagenomics research in developing countries is the limited capacity of tool development for genomics and bioinformatics approaches (Djikeng, Nelson et al. 2011). Although many computing grids exist, outsourcing analyses and computing is not always the best solution since data transfer requires significant bandwidth due to their size, constituting thus another bottleneck. Second of all, externalizing the analysis is not necessarily the best way to form local staff able to perform cutting-edge bioinformatics approaches and it might prove more efficient to create partnerships and regional centers for technology and resources.

Although, several success stories in building genomics initiatives in developing countries exist and could inspire metagenomics projects as well (Seguin, Hardy et al. 2008; Coloma and Harris 2009). Mexico has developed very effective translation of genomic findings into public health applications through a unique work plan called INMEGEN[7] (such as health promotion campaigns at Mexican sub-populations that might be at higher risk of certain chronic diseases). Another successful genomics initiative in Brazil where the Foundation for Research Support invested in projects relevant to the country and the rest of the developing world but are low on the list of priorities of the United States and Europe (e.g. sequencing the genes of the parasite that causes schistosomiasis that afflicts millions of people in Brazil). Previous successful examples of collaborations in genomics program could initiate new ones in the context of human microbiome studies (Djikeng, Nelson et al. 2011).

---

[7] Acronym for "Mexican National Institute of Genomic Medicine"

*Several metagenomics challenges for developing countries and pitfalls*

In spite of the success stories above, researchers in most developing countries do not have access to resources or the capacity to participate fully in genomics research (Coloma and Harris 2009). The ones that have access to local facilities are mostly those who were involved in the past decade in genome sequencing when they could build some sequencing capacities. An example of a successful collaboration is the joint effort of the International Livestock Research Institute (ILRI) in Nairobi and The Institute for Genome Research (Craig Venter Institute) to sequence and annotate the genome of a cattle parasite that causes important economic losses to small farmers in Africa and elsewhere. Beyond such "North–South" collaborations mainly aiming at capacity-building there is also a need for "South–South" collaborations. Access to training and capacity-building of human resources ought to be shared between developing countries fostering joint projects and mutualization of platforms.

Thanks to several funds, trusts and banks, there is increasing funding for research on diseases that affect the world's poor, such as the Bill & Melinda Gates Foundation (BMGF)[8], the United States National Institutes of Health[9], the United Kingdom Welcome Trust[10], the Asian Development Bank[11], etc. However, as mentioned by different authors, the success of the funded project depends on successful engagement with the intended beneficiaries and support from the local government; "*Recent research in developing countries, such as the abandoned trials in Cameroon and Cambodia of tenofovir as pre-exposure prophylaxis against HIV infection has shown that even in studies where ethical issues have been addressed, challenges related to community engagement (CE) can still undermine the research* (Djikeng, Nelson et al. 2011). Absence of community engagement, corruption, non returning of PhD students, absence of initiatives, there are still many pitfalls that may slow down the access of developed counties to the technology, resources, and capacity allowing their scientists to participate in metagenomics research.

---

[8] http://www.grandchallengesgh.org
[9] http://grants1.nih.gov/grants/index.cfm
[10] http://www.wellcome.ac.uk/funding
[11] http://www.adb.org

# Conclusion

The very young field of metagenomics is encountering a great success and is offering the possibility to explore in high-resolution places whose landscape we could only have imagined before. Seeing the outstanding diversity of the microbial world at the gene level is not enough to understand the functions and dynamics of the constituting parts. Ultimately, integrated analysis of metagenomes, metatranscriptomes, metaproteomes and meta-metabolomes will be needed to understand the microbial systems biology (Sleator, Shortall et al. 2008). Achieving such integration necessitates interdisciplinary efforts and continuous development of appropriate bioinformatics tools to decipher the complex biological networks underlying molecular, functional and community structure. The *in silico* investigation of biological networks could be quite effective in identifying central connected components that could bring, at a later time, more insight on their functionality and dynamics within the system.

International projects such as MetaHIT and HMP have already released a wealth of data on ecosystems along with the corresponding reference catalogues and their available functional and phylogenetic annotations. Making data public as well as bioinformatics/biostatistics tools is crucial in helping scientists from developing countries start adding their own research. Being part of the metagenomics race is of uttermost economical importance for developed but also for developing countries. Being able to improve human health as a result of metagenomics studies would not only help taking an important medical burden away from the society but also preserve active individuals that would participate to the economy. Many biotechnological applications still to be imagined could improve people's lives and reduce their impact on the environment. It is important for developing countries to "hop-on" the metagenomics train and join the international effort. North-South or South-South collaborations, legislation modification, sample trading, etc. are some of the strategies that could buy them the ticket. It is never too early to think big and invest in developing the needed ressources and start this very exciting and promising journey.